\documentclass[3p,times]{elsarticle}

\usepackage{ecrc}

\usepackage{epstopdf}

\volume{00}

\firstpage{1}

\journalname{Nuclear Physics A}

\runauth{V. Rolando \textit{et al}}

\jid{nupha}

\jnltitlelogo{Nuclear Physics A}

\usepackage{graphicx}
\usepackage{amsmath,amssymb}
\usepackage{amssymb}
\usepackage{amsthm}
\usepackage{subfig}

\usepackage{lineno}

\biboptions{sort&compress}

\usepackage{color}

\newcommand{\oneD}{$(1+1)-$D}
\newcommand{\twoD}{$(2+1)-$D}
\newcommand{\threeD}{$(3+1)-$D}
\newcommand{\echo}{ECHO-QGP}

\begin{document}

\begin{frontmatter}



\title{Heavy Ion Collision evolution modeling with ECHO-QGP}



\author[unife,infnfe]{\underline{V. Rolando}}
\author[unifi,infnfi]{G.~Inghirami}
\author[infnto]{A. Beraudo}
\author[unifi,infnfi,inaffi]{L. Del Zanna}
\author[unifi,infnfi]{F. Becattini}
\author[infnfi]{V. Chandra}
\author[infnto]{A. De Pace}
\author[infnto]{M. Nardi}

\address[unife]{Dipartimento di Fisica e Scienze della Terra, Universit\`a di Ferrara, Via Saragat 1, I-44100 Ferrara, Italy}
\address[infnfe]{INFN - Sezione di Ferrara, Via Saragat 1, I-44100 Ferrara, Italy}
\address[unifi]{Dipartimento di Fisica e Astronomia, Universit\`a di Firenze, Via G. Sansone 1, I-50019 Sesto F.no (Firenze), Italy}
\address[infnfi]{INFN - Sezione di Firenze, Via G. Sansone 1, I-50019 Sesto F.no (Firenze), Italy}
\address[infnto]{INFN - Sezione di Torino, Via P. Giuria 1, I-10125 Torino, Italy}
\address[inaffi]{INAF - Osservatorio Astrofisico di Arcetri, L.go E. Fermi 5, I-50125 Firenze, Italy}
\begin{abstract}
We present a numerical code modeling the evolution of the medium formed in
relativistic heavy ion collisions, ECHO-QGP. 
The code solves relativistic hydrodynamics in \threeD, 
with dissipative terms included within the framework of 
Israel-Stewart theory; it can work both in Minkowskian and in Bjorken 
coordinates. Initial conditions are provided through an implementation of the 
Glauber model (both Optical and Monte Carlo), while freezeout and particle 
generation are based on the Cooper-Frye prescription. 
The code is validated against several test problems and shows remarkable 
stability and accuracy with the combination of a 
conservative (shock-capturing) approach and the high-order methods employed. 
 In particular it beautifully agrees  with the semi-analytic 
solution known as Gubser flow, both in the ideal and in the viscous 
Israel-Stewart case, up to very large times and without any ad hoc
tuning of the algorithm.
\end{abstract}


\end{frontmatter}



\section{Introduction}
\label{sec:intro}
During the past years hydrodynamics has been aknowledeged as the most powerful 
tool to study the evolution of the medium formed in high-energy nuclear 
collisions.
For this reason a variety of calculations and numerical codes have been developed,
starting from the simple longitudinal boost-invariant Bjorken \oneD \:
calculation \cite{Bjorken:1982qr}, 
passing through \twoD \: and \threeD \: codes solving ideal hydrodynamics 
\cite{Kolb:2000sd,Kolb:2002ve,Kolb:2003dz},
up to more complex and structured codes simulating a full \threeD\: viscous 
evolution for the QGP fluid\cite{Luzum2009,Baier2006,Romatschke:2009im,
Ryu:2012at,Gale:2012rq,Schenke:2010nt}. \\
In order to solve dissipative hydrodynamics, one needs to deal with a set of 
non-linear partial differential equations, which includes the conservation
of the energy momentum tensor ($\partial_\mu T^{\mu\nu} = 0$)
and  of conserved charges like the baryon number ($\partial_\mu N^\mu = 0$), 
together with the time evolution of the bulk viscosity ($\Pi$) and the shear 
stress tensor ($\pi^{\mu\nu}$).
The system is closed by the choice of a suitable Equation of State (EoS).  
While the naive relativistic extension of the Navier-Stokes equations is 
affected by a well known causality problem, a consistent theoretical setup to 
include dissipative effects is the one formulated by Israel and Stewart 
\cite{Israel:1979wp} .

In this context \echo\: was presented last year \cite{DelZanna:2013eua},
equipped with second-order 
treatment of causal relativistic viscosity effects and unique features such as 
the possibility of choosing between two different metric tensors and of solving 
either purely ideal or viscous hydrodynamics equations.
As the other analogous tools, ECHO-QGP is characterized by a modular structure, 
allowing one to address the modeling of the initial conditions, the solution of 
the hydrodynamic evolution of the medium -- representing the core of the code -- 
and, eventually, the simulation of the final particle decoupling. 
\echo\: is highly customizable, allowing the user to choose among a variety of 
initial conditions (including both optical and Monte Carlo Glauber model) 
as well as the possibility of defining an energy density (or entropy density)
profile from scratch. The same principle applies to the equation of state: \echo\: 
can handle both analytical and tabulated equations of state. The decoupling stage 
exploits the Cooper-Frye prescription\cite{Cooper1974}, allowing the calculation 
of a mean spectrum of or the Monte Carlo generation of  a
discrete set of particles. The hypersurface 
detection has been recently upgraded with the embedding of a refined algorithm 
which creates a smooth mesh\cite{Huovinen:2012is}.

In order to ensure the reliability of the solving algorithm for the evolution 
stage, in our original publication \cite{DelZanna:2013eua} we 
showed how ECHO-QGP is able to overcome several numerical tests, displaying a 
beautiful agreement with some special analytic solutions. In the present 
contribution the above efforts in validationg the code will be briefly 
reminded and extended to the case of the so called Gubser-flow (both ideal and 
viscous), meanwhile appeared in the literature \cite{Gubser:2010ui,
Gubser:2010ze,Marrochio:2013wla}. The latter represents a 
very important test, since the code has to reproduce a very non-trivial flow in 
(2+1)D in the presence of non-vanishing viscosity and relaxation time. ECHO-QGP 
turns out to be able to overcome such a test.

Finally, having at our disposal such a validated tool, we will outline our 
future programs.

\section{Testing \echo} \label{sec:test}
\begin{figure}[!ht]
    \subfloat[Shock tube test: velocity profile \label{fig:shock_vx}]{%
      \includegraphics[width=0.45\textwidth]{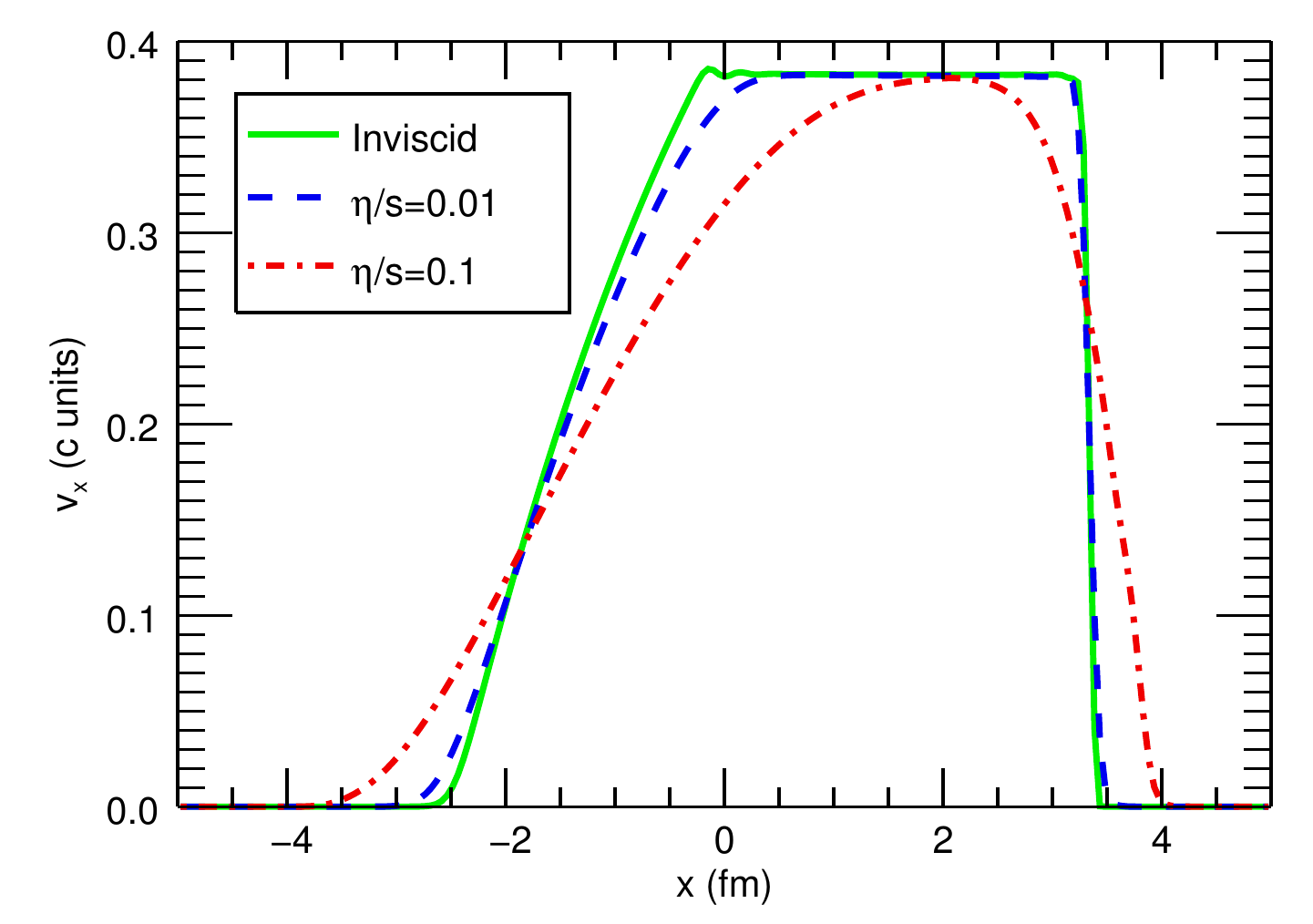}
    }
    \hfill
    \subfloat[Shock tube test: energy density profile \label{fig:shock_en}]{%
      \includegraphics[width=0.45\textwidth]{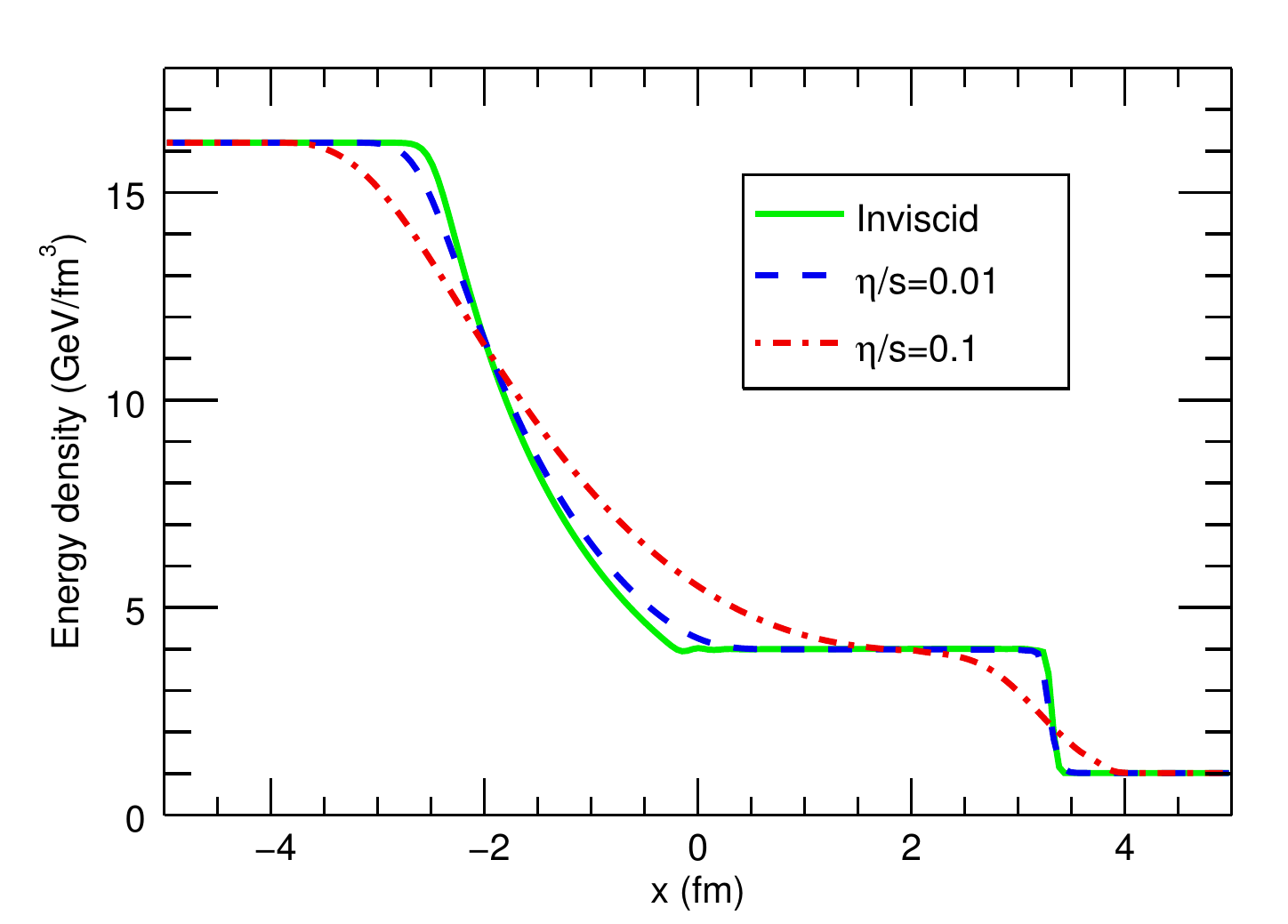}
    }
    \caption{Shock tube test for \echo.
    The test is performed placing an  initial diaphragm along the
    diagonal of a square box (201 points and size 10 fm along
    x and y) adopting Minkowskian Cartesian coordinates, and
    letting \echo\: evolve from t=1 up to t=4 fm$/$c. The temperatures 
    in the sides of the diaphragm are chosen to be
    $T^L = 0.4$ GeV and $T^R = 0.2 $ GeV.  It can be appreciated 
    how the viscosity affects the evolution, and how \echo\: handles shocks in 
    the ideal case without the introduction of dumps.
}
    \label{fig:shocktube}
  \end{figure}

ECHO-QGP has been widely tested, displaying agreement with analytic and 
semi-analytic solutions and other publicly available numerical codes.\\
Among the various test performed, an important role is played by the
 (2+1)-D shock-tube problem. Shock-capturing numerical schemes are 
 designed to handle and evolve discontinuous quantities invariably arising due 
 to the nonlinear nature of the fluid equations. 
In order to validate these codes, typical tests are the so-called 
shock-tube problems, performed in the  heavy-ion field 
also by \cite{Molnar:2009tx}. 
The test gives a good esteem of the behavior of the fluid in presence of 
non-vanishing $\eta/s$.
In figure \ref{fig:shocktube}, we compare the energy density 
profile for the viscid and unviscid case, where the high accuracy of the results 
and the absence of numerical spurious oscillations near the shock
front in the ideal case can be appreciated.

 Another fundamental qualifying test which numerical hydrodynamic codes have
 recently started addressing is the so called Gubser-flow. 
 The latter is a (2+1)D solution for a conformal fluid (with an 
 EoS $P=e/3$) characterized by longitudinal boost invariance and non-trivial 
 azymuthally symmetric radial expansion. It was derived first by
 Gubser \textit{et al.} \cite{Gubser:2010ze,Gubser:2010ui} for an ideal fluid
 and extended only last year to the viscous case, within the Israel-Stewart 
 setup, by Marrochio \textit{et al.}~\cite{Marrochio:2013wla}.
 The derivation of the solution is based on a rescaling of the metric (Weyl 
 rescaling) and on a change of coordinates, which allows one to exploit at best 
 the symmetries of the system arising from conformal invariance.
 In particular, one applies to the metric 
 $ds^2\!=\!-d\tau^2\!+\!dr^2\!+\!r^2d\phi^2\!+\!\tau^2d\eta^2 $ the rescaling 
 $ds^2 \rightarrow d\hat{s}^2\equiv ds^2/\tau^2$. 

\begin{figure}[!ht]
    \subfloat[\label{fig:gub_pixx}]{%
      \includegraphics[width=0.45\textwidth]{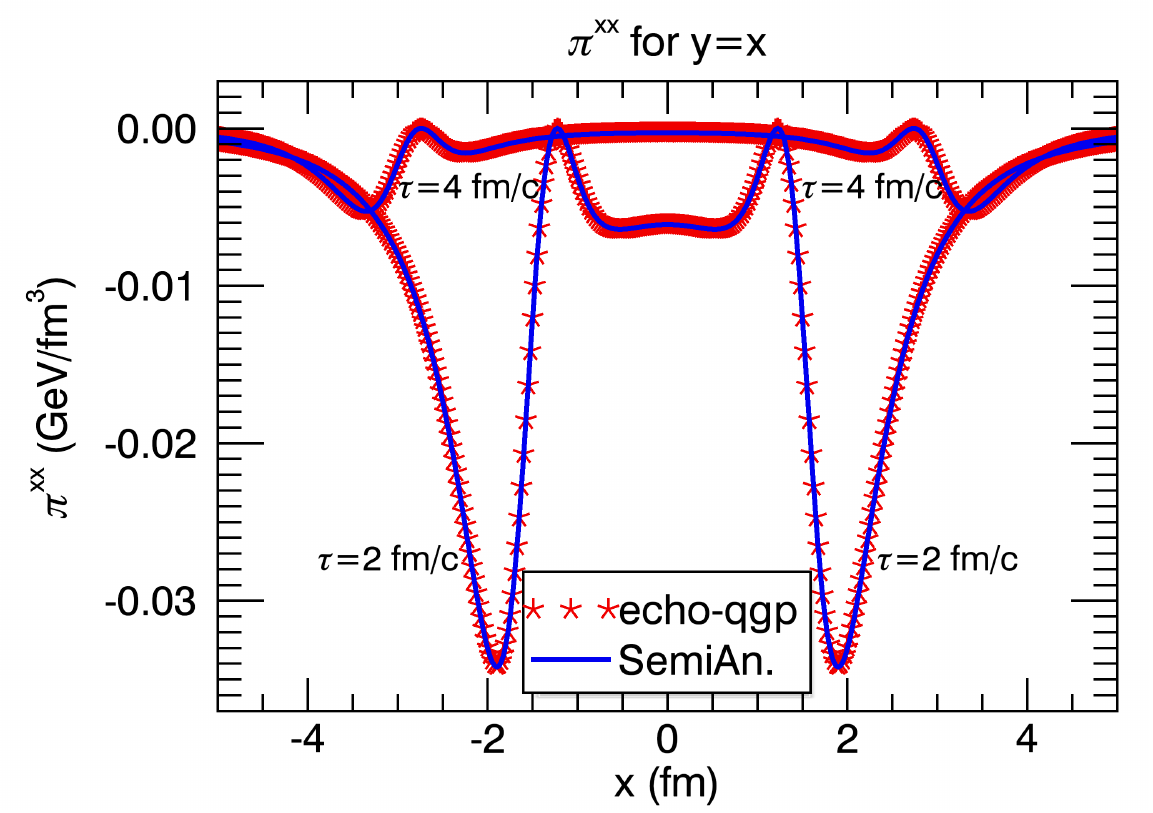}
    }
    \hfill
    \subfloat[\label{fig:gub_pixy}]{%
      \includegraphics[width=0.45\textwidth]{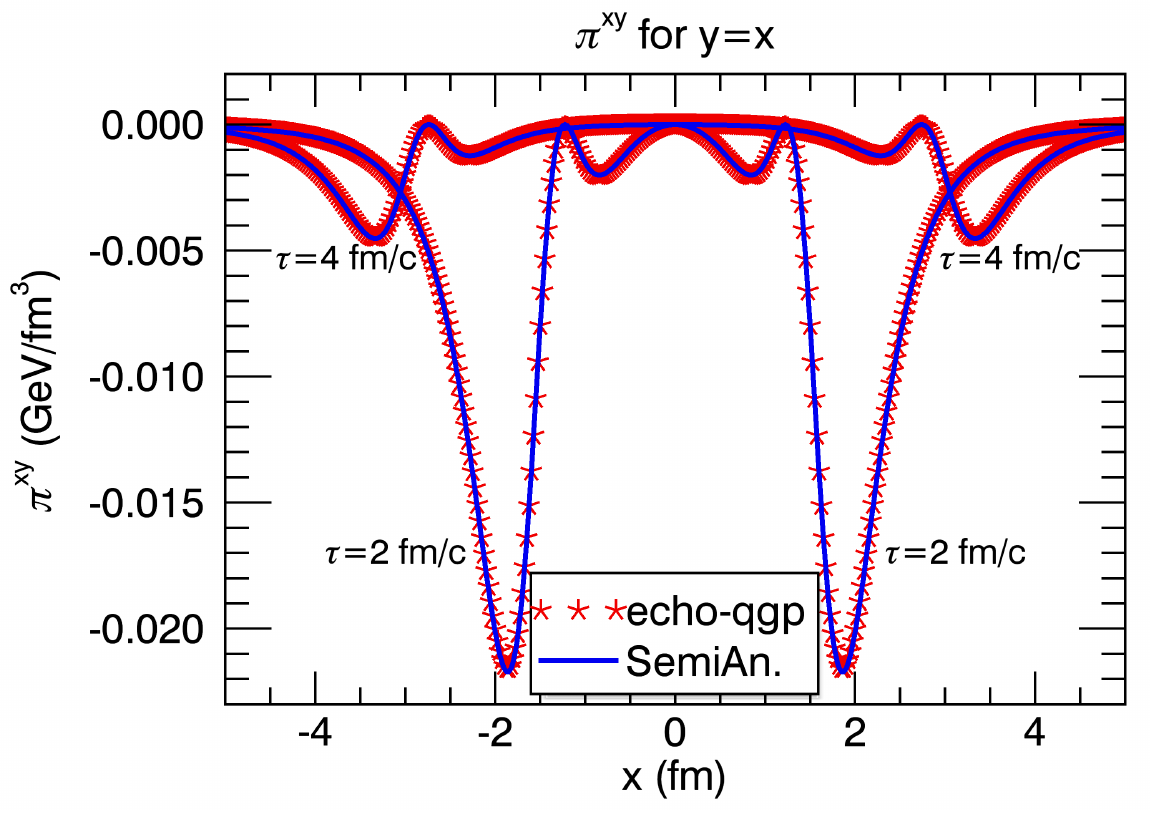}
    }\\
    \subfloat[\label{fig:gub_pizz}]{%
      \includegraphics[width=0.45\textwidth]{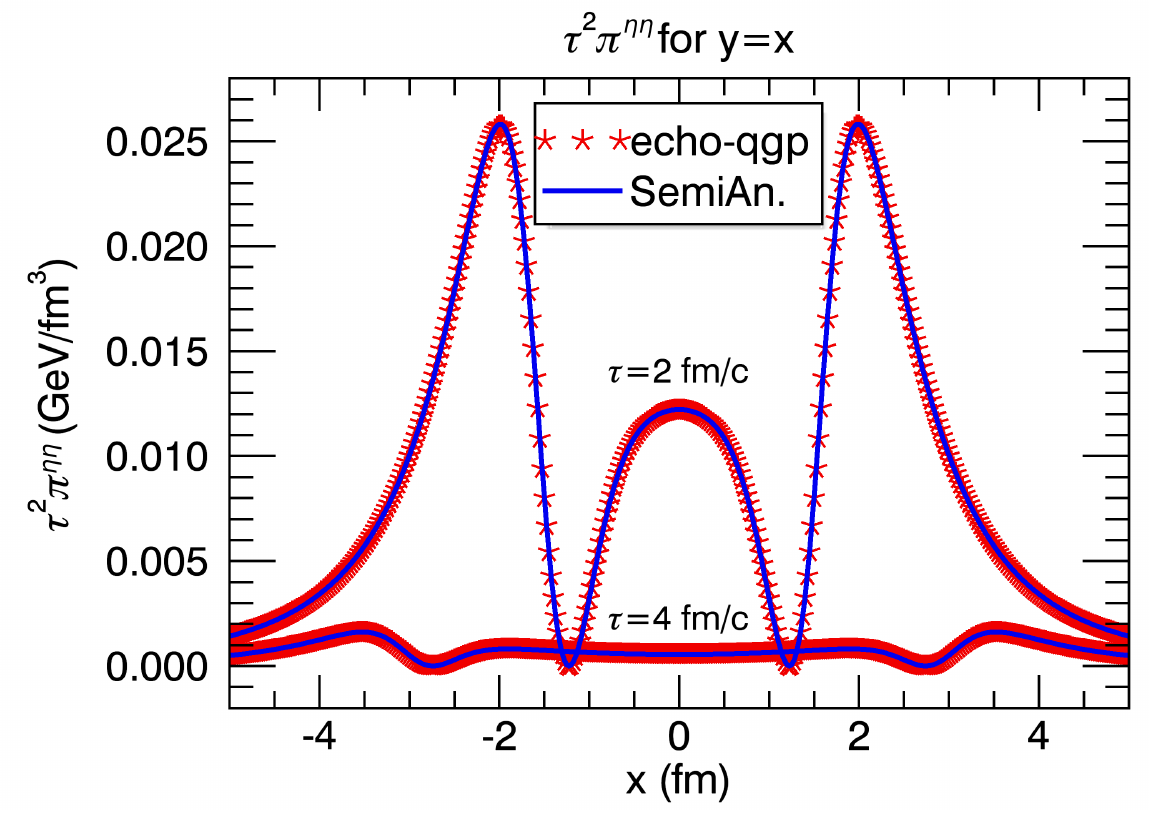}
    }
    \hfill
    \subfloat[\label{fig:gub_T}]{%
      \includegraphics[width=0.45\textwidth]{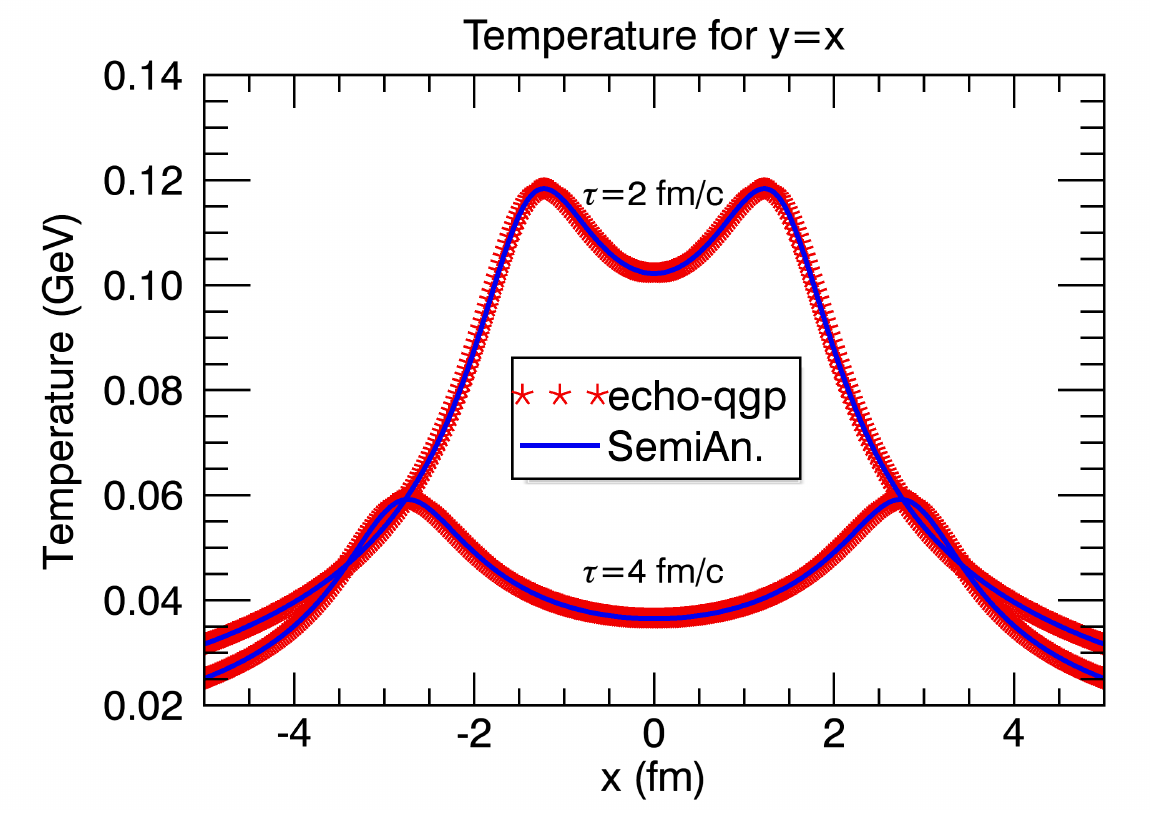}
    }\\    
    \caption{ \label{fig:gubserpanel} 
    Comparison between the semi-analytic solution, eq. \ref{eq:gubser}
    and an \echo\: \twoD \: simulation within the Bjorken coordinate metrics.
    The test is performed letting the fluid evolution 
    be initialized with a symmetric energy density profile together, 
    created ad hoc. The simulation has been performed with a grid of 0.025 fm in 
    space and 0.001 fm in time. The shear viscosity to entropy density ratio 
    is set to $\eta/s=0.2$, while the shear relaxation time 
    is $\tau_R = 5\eta/(e+P)$. 
    The energy scale is fixed at $q=1$ fm$^{-1}$\, while the
    equation of state must be the one of a 
    conformal fluid, we choose $e=3P$.\\
    The excellent agreement between the two can be appreciated for 
    every component of the shear stress  tensor (we show here just 
    $\pi^{xx}$, $\pi^{xy}$ and $\pi^{\eta\eta}$, but all the components present
    the same agreement), and for the thermodynamic variables (again we just show 
    the temperature). For the sake of clearness,
    we show here just two close  time steps ($\tau=2$ and $\tau=4$), 
    but the quality of the result is maintained up to much higher times. 
    We remark then \echo\: reproduces the semi-analytic solution without any 
    need of fine-tuning.}
  \end{figure}

 In addition, one has to perform the coordinate transformation 
 $d\hat{s}^2=-d\rho^2+\cosh^2\!\!\rho\,(d\theta^2+\sin^2\!\!\theta\, 
  d\phi^2)+d\eta^2,\label{eq:rescaledmetric}		$
 where the new coordinates are given by
  $\sinh\rho\!\equiv\!-\frac{1-q^2(\tau^2-r^2)}{2q\tau}$ 
 and 
  $\tan\theta\!\equiv\!\frac{2qr}{1+q^2(\tau^2-r^2)}$, 
 where $q$ is an arbitrary energy scale.
 In the new space (with quantities labeled by a hat) the fluid is at rest, 
  $\hat{u}_\mu\!=\!(-1,0,0,0)$, and the only equations to solve are:
  \begin{align}
  \frac{1}{\hat{T}}\frac{d\hat{T}}{d\rho}+\frac{2}{3}\tanh\rho
  =\frac{1}{3}\tanh\rho\,\bar{\pi}^\eta_\eta,
  &&
  \hat\tau_R\left[\frac{d\bar\pi^\eta_\eta}{d\rho}
  +\frac{4}{3}\left(\bar\pi^\eta_\eta\right)^2\tanh\rho\right]
  +\bar\pi^\eta_\eta=\frac{4}{3}\frac{\hat\eta}{\hat{s}\hat{T}}\tanh\rho,
  \label{eq:gubser}
  \end{align}
  where $\bar\pi\!\equiv\!\hat\pi/\hat{T}\hat{s}$. 
  Physical quantities in Minkowski space can be obtained from the above 
  equations through the mapping
  \begin{equation}
  T=\hat{T}/\tau,
  \qquad 
  u_\mu=  \tau\frac{\partial\hat x^\alpha}{\partial x^\mu}\hat{u}_\alpha,
  \qquad
  \pi_{\mu\nu}=
  \frac{1}{\tau^2}\frac{\partial\hat x^\alpha}{\partial x^\mu}
  \frac{\partial\hat x^\beta}{\partial x^\nu}\hat\pi_{\alpha\beta}
  \end{equation}
  and compared to the numerical solution provided by ECHO-QGP, as displayed 
  in Fig.~\ref{fig:gubserpanel}.

  The results have been obtained  without any fine tuning of the parameters or 
  modifications of the reconstruction algorithm. Nonetheless, \echo\:
  reproduces with very high accuracy (the discrepancy is on average of the order 
  of 0.1\%) and up to late times (we verified the agreement up to 10 fm/c)
  the temperature (see Fig. \ref{fig:gub_T}), all the shear-stress tensor 
  components (see Figs. \ref{fig:gub_pixx},\ref{fig:gub_pixy},\ref{fig:gub_pizz}) 
  and obviously the flow (not shown).
  
\section{Conclusions}
\echo\: 
guarantees
stability and precision, as well as completeness in
documentation and ease in the usage. We presented all those \echo\: features in 
\cite{DelZanna:2013eua} and we remark its suitablity to model heavy-ion 
collisions showing how well it reproduces the flow and shear stress tensor 
in the viscous relativistic frame proposed in ref. \cite{Gubser:2010ui,
Gubser:2010ze,Marrochio:2013wla}. 
Moreover, \echo\: has been positively used for the investigation  fluctuation
propagations\cite{Floerchinger:2013tya,Floerchinger:2014fta}, and it is currently 
used to study vorticity effects on the directed flow (in preparation).\\
\section{Aknowledgements} 
We would like to thank G. Denicol and the authors of the reference 
\cite{Marrochio:2013wla}
for interesting discussions and advices about the semi-analytic solution.
V.R. would like to thank prof. R. Tripiccione, G. Pagliara and A. Drago for 
their help in many interesting discussions about the decoupling stage.\\
This work has been supported by the Italian Ministry of Education
and Research, grant n. 2009WA4R8W.




\bibliographystyle{elsarticle-num}
\bibliography{rolando_QM14}







\end{document}